\documentclass[12pt, preprint]{aastex}
\def\sqig{$\sim$}
\def\degrees{$^{\circ}$}
\def\gcas{$\gamma$\thinspace Cas}
\def\1700{4U\thinspace 1700-37}

\def\src{4U2206+54}

\begin{document}

\title{\src\ - an Unusual High Mass X-ray Binary with a 9.6 Day Orbital
Period but No Strong Pulsations }

\author{Robin H.D. Corbet\altaffilmark{1, 2} \&
Andrew G. Peele\altaffilmark{3}}
\altaffiltext{1}{Laboratory for High Energy Astrophysics,
Code 662, NASA/Goddard Space Flight Center, Greenbelt, MD
20771; corbet@lheamail.gsfc.nasa.gov }
\altaffiltext{2}{Universities Space Research Association}
\altaffiltext{3}{University of Melbourne, School of Physics,
Parkville, VIC 3010, Australia}

\begin{abstract}
Rossi X-ray Timing Explorer All-Sky
Monitor observations of the X-ray source \src, previously
proposed to be a Be star system, show the X-ray
flux to be modulated with
a period of approximately 9.6 days. If the modulation is due to orbital variability then
this would be one of the shortest orbital periods known
for a Be star X-ray source. However, the X-ray luminosity
is relatively modest whereas a high luminosity would be predicted if the
system contains a neutron star accreting from the denser inner regions of
a Be star envelope. Although a 392s pulse period was previously reported
from EXOSAT observations, a reexamination of the EXOSAT light curves does
not show this or any other periodicity. An analysis of archival RXTE
Proportional Counter Array observations also fails to show any X-ray
pulsations.
We consider possible models that may explain the properties
of this source including a neutron star with accretion halted at the magnetosphere and an accreting white
dwarf.

\end{abstract}

\section{Introduction}

The X-ray source \src\ was discovered with Uhuru (Giacconi et al. 1972)
and subsequently the source position was precisely localized with HEAO
1 (Steiner et al. 1984). This led to the identification by Steiner
et al. of a V = 9.85 optical counterpart, BD +53\degrees 2790, which
these authors classed as a Be star based on its double-peaked H$\alpha$
emission line and derived a spectral classification of B1$\pm$1 from broad
band optical photometry.  A long term X-ray light curve of \src\ was also
obtained by Steiner et al. from the Ariel 5 Sky Survey Instrument (SSI) which,
although subject to large error bars, showed considerable variability. For
the distance of 2.5 $\pm$ 1 kpc derived by Steiner et al. (1984) from
optical photometry, the mean flux seen with Ariel V corresponds to a
luminosity of 7.0 $\times$ 10$^{34}$ ergs s$^{-1}$.  Saraswat \& Apparao
(1992; hereafter SA92) observed \src\ with EXOSAT three times with the
observations separated by about one year. The EXOSAT light curves showed
the source to be very variable and SA92 reported the possible detection
of 392s pulsations. Saraswat et al. also obtained Rossi X-ray Timing
Explorer Proportional Counter Array (RXTE PCA) observations of \src\ on
1997 March 11 and 13 which had not been published.  Additional optical
observations are reported in Corbet (1986a) which showed the source
to continue to exhibit strong double peaked H$\alpha$ emission and IUE
ultraviolet spectra are reported by Teodorani et al. (1994).

This source thus initially appeared to be a fairly typical low luminosity Be/neutron
star binary with a relatively long pulse period.
Be/neutron star binaries contain a non-supergiant late O or early B type
primary which possesses a variable equatorial disk from which
H$\alpha$ emission occurs. The neutron star, which is
often in an eccentric orbit, may accrete from
this disk when it is present.
Transient variability is typically observed in these systems
due to both orbital modulation of the accretion rate, if the
orbit is significantly eccentric, and
variations in the Be star's disk (see e.g. Stella, White \& Rosner
1986, Bildsten et al. 1997, and references therein).  
The low luminosity and proposed long pulse period
of \src\
suggested that the source may have a relatively long orbital period of
perhaps hundreds of days. We report here on a search for
the orbital period of \src\ using Rossi X-ray Timing Explorer All Sky Monitor (RXTE ASM) data and
the discovery of a much shorter orbital period than expected of 9.6 days.
This surprisingly short orbital period 
prompted us to analyze the archival EXOSAT and RXTE PCA light curves
to search for the pulsations
reported by SA92
and we find no evidence for these. The short orbital period and lack of
pulsations now show \src\ to be much more unusual than previously thought.
In addition, recent reports of more extensive optical and ultraviolet
observations (Negueruela \& Reig 2001, Negueruela et al. 2001) appear
to call into question the Be classification of the primary star.
These authors argue that the primary star in this system is
not a Be star but instead is a very peculiar late O-type active star
and that the ultraviolet and blue spectrum is dominated by a star of
approximate spectral type O9.5V. This spectral classification is close
to a classification of O9.5III?p given prior to the discovery
of \src\ by
Hiltner (1956) which Steiner et al. (1984) were apparently unaware of.
This revised spectral type yields
a somewhat larger source distance of \sqig3 kpc and increases the implied
mean source flux seen to about 10$^{35}$ ergs s$^{-1}$.

The RXTE ASM observations are presented in Section 2, the
search for pulsations from pointed EXOSAT and RXTE data is described
in Section 3, possible models that may explain the properties of \src\
are given in Section 4, and Section 5 concludes.

\section{RXTE ASM Observations of \src}
The ASM (Levine et al. 1996) on board RXTE consists of three Scanning
Shadow Cameras, sensitive to X-rays in an energy band of approximately
2-12 keV, which perform sets of 90 second pointed observations (``dwells")
so as to cover \sqig80\% of the sky every \sqig90 minutes.  The ASM
provides light curves of several hundred sources including \src\ for
which light curves are routinely constructed.  The analysis presented here
makes use of both the background-subtracted light curve obtained from the
individual dwells and a daily averaged light curve constructed from the
flux measured in individual dwells. The Crab produces approximately 75
counts/s in the ASM over the entire energy range and ASM observations
of blank field regions away from the Galactic center suggest that
background subtraction is accurate to a systematic level about 0.1
counts/s (Remillard \& Levine 1997).

The daily averaged ASM light curve of \src, heavily smoothed in order
to make long term variability easier to see in this faint source, is
shown in Fig. 1.  The source is seen to be variable but is detected most
of the time during the approximately 5 years of coverage.  In order to
search for periodic modulation, and hence determine the orbital period,
we analyzed the light curve by calculating the power spectrum of the
unsmoothed dwell data and found, in addition to low frequency variability,
highly significant modulation on a period of \sqig9.57 days (Fig. 2).
In order to measure the period more precisely, and determine the error
on this, we fitted a sine wave to the light curve. In order to take
into account uncertainties caused by source variability as well as
instrumental uncertainties in flux measurement we scaled the errors on
individual points by a factor of \sqig1.7 to force the reduced $\chi^2$
of the fit to be 1.  From this fit we find a period of 9.568 $\pm$ 0.004
days and, from this fit and an examination of and the folded light curve
(shown in Fig. 3) we determine the epoch of maximum X-ray flux to be MJD
51006.1 $\pm$ 0.2. The folded light curve plotted in Fig. 3 has a profile
which is approximately sinusoidal. We note that this is accumulated over a
range of flux levels and it may be possible that profile changes could be
connected with overall flux or that other profile changes may occur. The
error bars in Fig. 3 are based on the distribution of fluxes in each
phase bin and thus have contributions from both statistical uncertainty
in the flux measurements and also changes in the overall flux level and
any profile changes.  In Fig. 4 we show a detail of the light curve in
a form less heavily smoothed than Fig. 1 with times of expected maxima
marked. This plot demonstrates that, when the source is relatively bright,
individual cycles can be directly seen in the ASM light curve.

By analogy with other high mass X-ray binaries,
the highly coherent period we find is expected to be
the orbital period. For example, recurrent outbursts,
particularly in Be star systems, can often occur
on the orbital period  (e.g. Stella, White \& Rosner 1986, Bildsten et al. 1997).
While the discovery of an orbital modulation was not unexpected, the
length of the period at 9.57 days is surprising for several reasons: (i)
the 392s pulse period reported by Saraswat would imply an orbital period
of a few hundred days if the system had parameters consistent with the
general correlation between pulse and orbital periods exhibited by most
Be/neutron star systems (Corbet 1986b); (ii) the X-ray flux of \src\
is relatively low at \sqig 10$^{35}$ ergs s$^{-1}$ - this would
again suggest that the system has a long orbital period as a neutron
star with a short orbital period would be able to accrete from the
denser inner regions of the Be star envelope; (iii) an orbital period
of this length
would be shortest known for a Be/neutron star system apart from a possible 0.7
or 1.4 day period reported from optical observations of the SMC source
AX J0051-73.3 (Cook 1998, Coe \& Orosz 2000).  The shortest known X-ray
period for a Be/neutron star system is the 16.65 day period
of A0538-66
(Skinner et al. 1980). A0538-66 has a super-Eddington luminosity when
at its brightest and shows extreme optical spectroscopic and photometric
variability (e.g. Charles et al. 1983, Corbet et al. 1985) unlike \src\
which has not so far exhibited such dramatic behavior.

\section{A Search for Pulsations}

\subsection{Archival EXOSAT Observations}
Because of the unexpectedly short orbital period that we found in the ASM data, we
decided to reexamine the possible X-ray pulsations reported by SA92.
We note that a fundamental limitation with the SA92
analysis is that they used a procedure not well suited to a search for
coherent periodic pulses (pulsations). Instead of Fourier analyzing the entire
light curve, SA92 divided their light curves into 5 segments of 2000
seconds each, computed a power spectrum of each segment, smoothed it,
and then added the power spectra.  This procedure is not ideal because it
does not exploit the high coherence that a truly periodic pulse would have for
the entire observation duration. In addition, the smoothing of a power
spectrum is inappropriate in a search for truly periodic modulation.
This type of procedure is, however, useful in the search for either
quasi-periodic oscillations (QPOs) or for cases where the pulse loses coherence
during an observation due to, for example, pulse period changes caused
by accretion torques or orbital Doppler shifts.  For \src, as
with other Be star systems, a pulse period of several hundred seconds
is unlikely to be smeared out beyond the point of detectability
by orbital motion during
the course of an observation lasting less than one day.

EXOSAT observed \src\ three times: on 1983 August 8, 1984 December 7,
and 1985 June 27.
Using our orbital ephemeris we estimate that these observations
were centered on orbital phases of
0.28 $\pm$ 0.20, 0.20 $\pm$ 0.18, and 0.30 $\pm$ 0.18 respectively.
The source was strongly detected in the 1983
and 1985 observations but not in 1984. We therefore obtained
the background subtracted 0.8 to 8.9 keV  
background subtracted 
light curves from the HEASARC for the 1983 and 1985 observations.
The 1983 light curve has 10 s time resolution and the 1985 light
curve has 1 s resolution. The last \sqig25\% of the 1985 light
curve as shown in Fig. 3 of SA92, 
which contained some flaring activity, was not available in the
HEASARC archive. Using these archival light curves
we calculated power spectra, constructing a single power spectrum
for each of the observations and without performing any smoothing. 
These resulting power spectra
are shown in Fig. 4
and we find no evidence for any
pulsations for periods longer than 2 seconds. Further,
we see no signal with an amplitude
greater than  \sqig 10\% and \sqig7.5\% 
at the period of about 392s reported by SA92 from
the 1983 and 1985 observations respectively.

\subsection{Archival RXTE PCA Observations}
In addition to the EXOSAT light curves of \src\ 
observations were also obtained by Saraswat et al. using
RXTE's main detector,
the Proportional Counter Array (PCA).  This instrument is described in
detail by Jahoda et al. (1996). It is an array of five Proportional Counter
Units (PCUs)
that have a total effective area \sqig 6500 cm$^2$ and has a band pass
of approximately 2 -- 60 keV.
Due to instrumental problems not all PCUs are always in operation
and the observations reported here use varying numbers of PCUs.
The PCA has a collimator which gives a
field of view of 1\degrees\ full width half maximum.

The RXTE PCA observations of \src\ were made on 1997 March 11 (04:14 to 09:46;
``Observation 1'') and 13 (06:02 to 08:46; ``Observation 2'') with a total exposure time
of 20ks. The PCA data collection modes were ``Good Xenon'', which provide
full spectral information with a time resolution of about 1 $\mu$s, in addition to
the two standard modes: ``Standard 1'' and ``Standard 2''.  ``Standard 2''
provides the full spectral information but at a time resolution of only 16s.
The light curves from these observations are shown in Fig. 5. It can be
seen that \src\ was significantly brighter in Observation 1 and we note
that fortuitously the first observation was made close to the phase
of the orbital maximum ($\phi$\sqig0.02) while the second observation
was made at $\phi$\sqig0.23.

In order to search for any pulsations in the RXTE light curves we
employed two techniques. First, to search for any slow pulsations, 
we
employed the same technique that we used for
the EXOSAT light curves on the Standard Mode 2 light curves. We calculated
power spectra for the two observations independently and the resulting power spectra
are shown in Figure 6. 
Again we find no evidence for any pulsations, in particular none are
found at the 392 s period reported by SA92.  Next, in order to search
for any higher-frequency pulsations, we calculated a power spectrum
from the background subtracted ``Good Xenon'' light curve. The light
curve was binned into 2$^{-10}$s (i.e. approximately
1 ms) bins, and then power spectra
were calculated for stretches of 256s and these power spectra were then
summed. Again no evidence is found for any pulsations and no QPOs were
found. We note that Negueruela \& Reig (2001) in an independent analysis
of a portion of this data set also reach the conclusion that no low
frequency pulsations are present.

We fitted the PCA spectra of \src\ from the two observations
and find that a typical X-ray
pulsar spectrum of an absorbed power-law with a high energy cutoff
(e.g. White, Holt, \& Swank 1983) gives a good fit and the parameters of
the fits are given in Table 1. We note that the decrease in flux found
in the second observation was
apparently not due to an increase in the absorbing column density

\section{Discussion - the Nature of \src}
\src\ is clearly unusual for a Be star X-ray source
with its short orbital period,
low luminosity, and apparent lack of pulsations.
In this section we discuss some possible models
that may account for the properties of this system. Some of these
models may account on their own for all of the properties of this
system and some may account for just some of the source properties.

\subsection{Be/Neutron Star Binary}

Despite the peculiarities in the optical/UV spectrum  of \src\ it
still appears to meet the basic definition of a ``Be'' or ``Oe'' star
as a main sequence star of early spectral type with variable H$\alpha$
emission. We may compare how the observed X-ray properties, in particular
the orbital modulation, fit with a Be/neutron star interpretation.
In high mass X-ray binaries orbital modulation may arise in several
ways. In those systems which have short orbital periods and a supergiant
primary eclipses may occur. However, in Be/neutron star binaries no eclipses
of the X-ray emitting region
have been detected - the smaller radius of the primary and longer
orbital periods make the probability of the system inclination
being sufficiently high for eclipses to be seen considerably smaller.
In the case of \src, although it appears to have an orbital period shorter
than the majority of Be star systems, the probability of the inclination
angle being high enough for an eclipse to occur is still low and
any eclipse would be brief. For a 1.4 M$_\sun$ neutron
star and \sqig10M$_\sun$ primary the orbital separation would be \sqig 300
light seconds but the radius of a \sqig B0V star is only \sqig 20 light seconds
(Allen 1973).
For these parameters the inclination angle would need to be
greater than \sqig 87\degrees\ for an eclipse to be seen,
and even an inclination angle of 90\degrees\ would yield a total eclipse duration of only \sqig 5 hours.
The optical spectrum of \src\ shows a strongly double peaked H$\alpha$
emission line. This double peaked structure suggests that we may be
viewing the source close to the plane of a circumstellar disk around
the primary star and that the orbital inclination may also be high if the disk
and orbital planes coincide.
However, the inclination angle may still not be
large enough for an eclipse to occur. 

Be star
systems often have eccentric orbits and, in these systems, the varying
orbital separation typically leads to variations in the mass-transfer rate. The
resulting modulation of the observed X-ray flux can be significantly
amplified as, at lower mass-transfer rates, accretion can be halted
almost completely by a magnetospheric barrier (e.g. Stella et al. 1986).
Another form of orbital modulation may occur
if the orbital plane is inclined from the plane of the circumstellar disk
around the Be star. In this
model enhanced accretion
may occur twice per orbit (e.g. Shibazaki 1982, Priedhorsky \& Holt 1987).
This orbital
modulation mechanism could apply even for circular orbits and again can
be amplified by a sudden overcoming of the magnetic barrier. 
Disk/orbital-plane offset has been suspected for several systems and the
best evidence to date for such a process to be at work may come from
XTE J1946+274 (Wilson, Finger, \& Scott 2000). 
We examined the ASM light curve folded on twice
our proposed orbital period and there were no large differences
between ``odd'' and ``even'' maxima or minima.
If the orbital system is actually twice the 9.57 day
period then the system would be less unusual.
However, even if the orbital period
of \src\ is \sqig20 days we still lack strong evidence
for the pulsations that would be expected if the
compact object in the system is an accreting neutron
star. 
If the system is actually pulsing but at a level below
our detection threshold then many of
the peculiarities might disappear.
While accretion onto a black hole would produce X-ray emission without
pulsations being present, the X-ray spectrum of 
\src\ appears typical for a neutron star.

\subsection{Accreting White Dwarf}

There are a large number of systems known that consist of a neutron star
accreting from mass lost from an early type companion. 
In principle
it should be possible for similar systems to exist where the accreting
object is a white dwarf rather than a neutron star (e.g. Waters et
al. 1989).
For low mass X-ray binaries the white dwarf counterparts are cataclysmic
variables which are systems consisting of a white dwarf accreting from
a Roche lobe filling late type companion.
However, the high mass white
dwarf equivalents have not yet been definitely shown to exist.
High mass systems with an accreting white dwarf would have a lower luminosity than an equivalent
neutron star system due to the shallower potential well of the white
dwarf. The possibility that \src\
might contain an accreting white
dwarf was proposed by SA92.  Accretion onto a white dwarf
could produce X-ray pulsations if the white dwarf is itself strongly
magnetized. However, in cataclysmic variables accreting non-magnetic white dwarfs are also known
where there is no modulation at the rotation period of the white-dwarf
(e.g. C\'ordova 1995).

Although no high-mass accreting white dwarf systems have yet been
definitely shown to exist, we note that
one object that has been proposed by some authors to
be such a system is \gcas.
This is a Be star X-ray source which
has a very low X-ray luminosity (\sqig10$^{33}$ ergs s$^{-1}$)
and has no clear evidence of pulsations, despite
occasional reports to the contrary which have been 
due to the large non-periodic variations present in the light curve
(e.g. Parmar et al. 1993 and references therein).
The nature of \gcas\ is still controversial and
several models have been proposed.
Kubo et al. (1989)
have
ascribed the X-ray emission from \gcas\ to accretion onto
a white dwarf. These authors base their argument for this on the basis
of the X-ray luminosity and the X-ray spectrum. 
However, other authors such as Robinson \& Smith (2000, and references
therein) have proposed non-binary models.
For \gcas\ no
strong evidence for it being a binary such as orbital
modulation of the X-ray flux has been found which may argue against
the white dwarf model.
In contrast,
the \sqig10 day modulation in the X-ray flux from \src\ 
which we have now measured appears to be strong evidence for binarity.
However, even though the luminosity of \src\ is
relatively low, it may still be too high to be
accounted for by accretion onto a white dwarf
(cf. Waters et al. 1989, Apparao 1991).

\subsection{Energy Release at a Neutron Star Magnetosphere}

A model that can produce X-ray luminosities intermediate
between accretion onto a neutron star and
onto a white dwarf is if accretion is halted,
and energy released at, the magnetosphere of
a neutron star.
The radius of the magnetosphere is determined by the accretion
rate and magnetic field of the neutron star. If the magnetospheric
radius is greater than the corotation radius, the radius at which
material corotating with the neutron star would be at its Keplerian
velocity, then the magnetosphere acts as a barrier to accretion onto
the surface of the neutron star. The interaction between the accreted
matter and and the magnetosphere will cause the rotation rate of the
neutron star to slow and so gradually increase the size of the
corotation radius. 
This model was initially suggested as a possible explanation
for low luminosity states in transient low mass X-ray binaries
(Stella et al. 1994).
It has also been proposed, for
example, for \gcas\ (Corbet 1996)
and was also invoked as a possible explanation
of a low intensity state observed in the Be/neutron star
binary A0538-66
(Campana et al. 1995, Corbet et al. 1997) where the average
X-ray luminosity was \sqig5.5$\times$10$^{36}$ erg s$^{-1}$,
significantly less than the super-Eddington luminosity exhibited
by this source in its brightest states. During this low intensity state
of A0538-66 no X-ray pulsations were detected although an
earlier bright state observation
had shown the presence of 69 ms pulsations (Skinner et al. 1982).
In Be/neutron star systems the very large orbital modulation
that can be present
is often attributable to the magnetospheric barrier
that turns on and off as the neutron star moves
around the Be star in an eccentric orbit and the accretion
rate varies and thus modulates the magnetospheric radius.
If this barrier is always closed
in \src\ then substantially less orbital modulation should
be observed although, if the orbit is significantly eccentric,
some modulation should still be observed as the accretion rate will
still depend on the distance from the primary.
In addition, as material is strongly inhibited from falling onto the neutron star
poles, strong pulsations would not be observed.
This model can thus explain both a lack of strong pulsations and a low
X-ray luminosity for \src. In addition, the source may be expected
to be active for a larger fraction of the time than ``standard'' Be/neutron
star binaries which effectively turn off when the magnetospheric
barrier closes. This would then be consistent with the long term
ASM light curve (Fig. 3) and the Ariel 5 SSI light curve (Steiner
et al. 1984) which
show \src\ to be detectable for a large fraction of the time.

If the magnetospheric ``accretion'' model is correct then it may still be
possible for some accretion onto a neutron star surface to be occurring
via material ``leaking'' onto the poles similar to that
observed in A0535+26 during a low state (Negueruela et al. 2000).
This would then produce modulation with a low pulsed fraction.
Given the short orbital period of \src\ a relatively short pulse period
could perhaps be expected based on the orbital/pulse period correlation
for Be/neutron star binaries (Corbet 1986).
Conversely, if accretion onto the surface
of a neutron star is taking place,
instead of being halted at a magnetosphere,
then, based on the relatively low X-ray flux,
a long
pulse period would be expected (Stella et al. 1986).

\subsection{Neutron Star Spin/Magnetic Axis Alignment}

In order to explain the lack of X-ray pulsations from \src\ we
can also make a comparison with another source which lacks pulsations.
This source is the bright high mass X-ray binary \1700\ which has an O6f supergiant
primary and a 3.4 day orbital period (e.g. Rubin et al. 1996
and references therein).
In the case of \1700\ accretion is thought to take place from
a strong stellar wind and, although its X-ray spectrum is similar
to those of X-ray pulsars, no pulsations have
been found from \1700: White et al. (1983) give
a 2\% upper limit on pulsations between 160 ms and 480s
and Gottwald report limits of 0.5 to 50\% for periods
between 3.9 ms to 50,000s.
For \1700\ it has been speculated that the magnetic and
rotation axes of an accreting neutron star in
the system may be co-aligned (e.g. White et al. 1983) which
would result in a lack of X-ray pulsations.
This model might plausibly also apply to \src. However, even though
this model can explain a lack of X-ray pulsations,
on its own it does not predict that the X-ray luminosity would
be unusually low.

If \src\ contains a neutron star with a very low magnetic field then
pulsations would also not occur. However, the high-energy cut-off
in the X-ray spectrum can be interpreted as being the sign of an
electron cyclotron
scattering feature which, if present, would imply a very strong magnetic field (see e.g. Chanmugam 1992).

\subsection{Wind Accretion}

Another result that demonstrates the peculiar properties of \src\
has been reported by Negueruela \& Reig (2001) and Negueruela et
al. (2001) who find that the optical and ultraviolet spectrum
is that of a very unusual late O type star.
A
major argument for this interpretation comes from a spectrum obtained with
the International Ultraviolet Explorer which Negueruela \& Reig (2001)
report as showing a clear O type spectrum and the overall optical/UV
spectrum shows signs of being composite.  Accretion in this system is
proposed by these authors to occur by accretion from a stellar wind rather than a Be
star type circumstellar envelope. This model could thus account for
the lower than expected X-ray luminosity as, since the primary
is not a supergiant, the wind should be relatively weak. 
However, this wind accretion does not, in
itself, lead to an immediate explanation for the lack of pulsations
and an additional mechanism such as spin/magnetic axis alignment may
need to be invoked.
The long
timescale changes observed in the X-ray flux would also appear to require
that the wind itself would be variable on these timescales in this model.
If the H$\alpha$ emission is not coming from a Be type disk,
then another site is required for it.
We note the possibility that some other peculiarity of the accretion 
process might result in peculiarities in the
optical spectrum.
For example, in the case where accreting material is halted 
at a neutron star magnetosphere
this material may be accelerated and expelled from the
system
in some form of ``wind''. The acceleration that will be experienced
at the magnetosphere will depend on the radius of the magnetosphere
and hence the accretion rate.  

If the primary star does posses a significant wind then this
may provide another mechanism for orbital modulation of the X-ray
flux. At
least for systems with a supergiant primary, orbital
modulation may occur due to variable absorption in the
line of sight caused by extended structures such as accretion wakes 
(e.g. Kaper, Hammerschlag-Hensberge, \& Zuiderwijk 1994
and references therein).

\section{Conclusion}

\src\ has several unusual properties, these include the lack of detectable
X-ray pulsations, an orbital period that would be short for a Be-type
X-ray binary, and a composite optical/UV spectrum.
The strongest candidates for models currently appear to be either (i)
accretion onto a neutron star magnetosphere from a Be star type envelope
or (ii) accretion onto a neutron star with aligned magnetic and  spin
axes from the wind of the O type primary. 
The critical difference between these models is the primary mode of
mass loss from the primary star.
Of course, some other, as yet
unexpected, phenomenon could be responsible for the unusual properties
of \src.

In order to understand the nature of \src\ additional observations
are required. 
Optical spectroscopy over an orbital cycle
might be capable of measuring radial velocity
changes and thus determining the mass of the compact object thought to be
present in the system.  High resolution X-ray spectroscopy of the 6.4/6.7 keV
iron line region may also be a possible discriminator between accretion
onto the surface of a
white dwarf or a neutron star (cf. Kubo et al. 1998 and
references therein).
A high statistical quality high-energy X-ray spectrum would be useful
to search for the presence of a cyclotron line. If this were to be found
this would be good evidence for the presence of a neutron star with
a strong magnetic field.
If this
source is currently accreting but with the accreted matter halted at a neutron
star's
magnetosphere then, if the accretion rate increases above the critical
value, or if the neutron star spins down sufficiently to make
the corotation radius larger than the magnetospheric radius, then the
barrier would open and material would accrete directly
onto the surface of the
neutron star itself. If this happened the source would then 
become dramatically brighter and exhibit X-ray pulsations.
Even if such
a large brightness change does not occur, a deeper search for the presence
of X-ray pulsations from \src\ may be useful. These may be present at
a very low level even if most accretion is halted by a magnetospheric
barrier.  X-ray spectroscopy at different orbital
phases could assist in understanding the cause of the orbital modulation,
if structures such as accretion wakes are present then they could be
revealed by systematic changes in the absorption column density.
The long term variations in the X-ray flux should be caused by changes
in the circumstellar environment from which the compact object is
accreting. Simultaneous long term optical spectrometry and photometry
may aid in determining whether this is a change in a Be star type
circumstellar envelope or a wind.

\acknowledgments
We thank Ignacio Negueruela for communication of results prior
to publication and the referee, Myron Smith, for useful comments.
This paper made use of data obtained through the HEASARC online service,
provided by the NASA/GSFC.

\pagebreak
\noindent
{\large\bf Figure Captions}

\figcaption[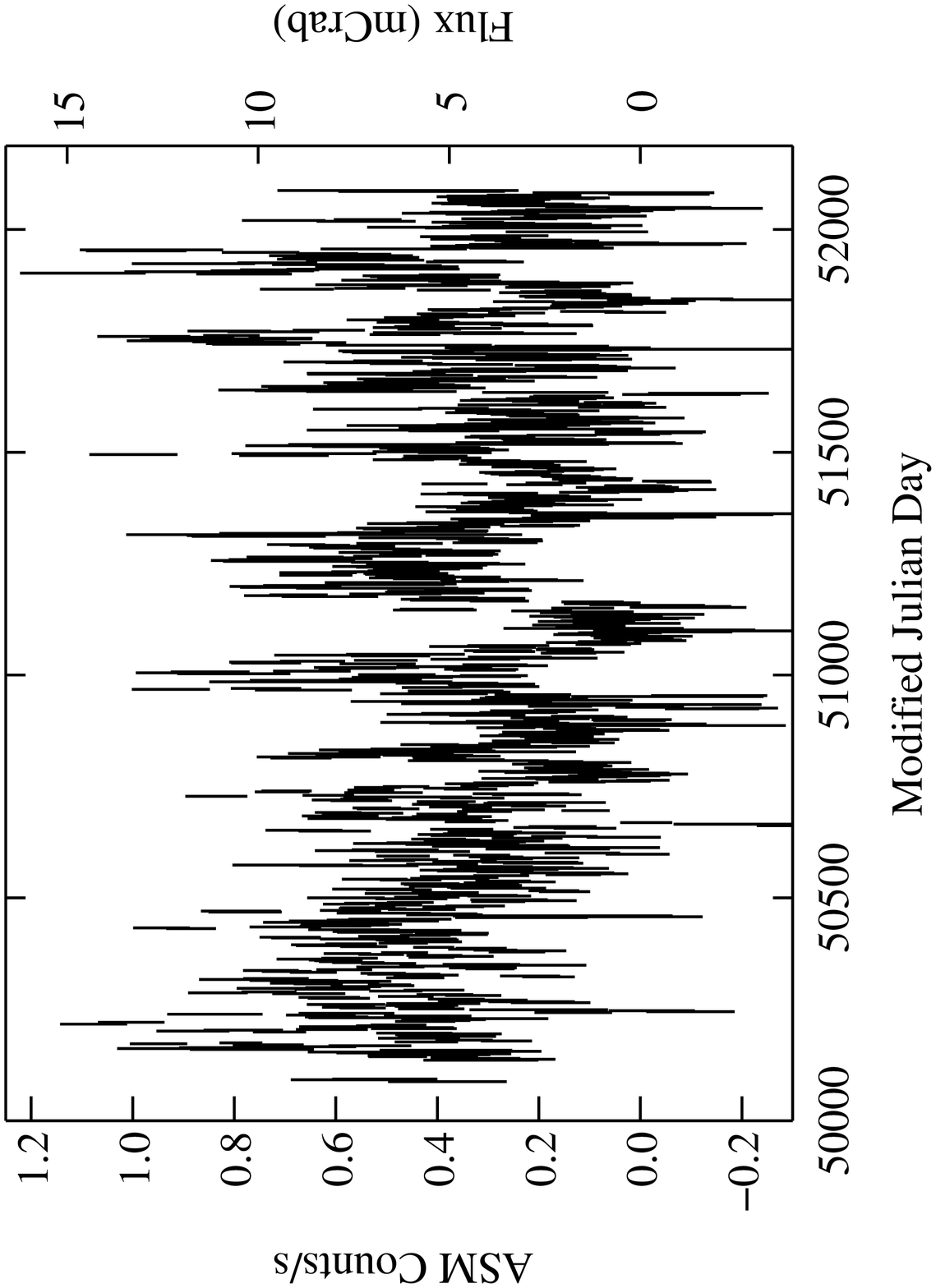]{The ASM light curve of \src. This is a smoothed and
then rebinned by a factor of two version of the one day light curve.}

\figcaption[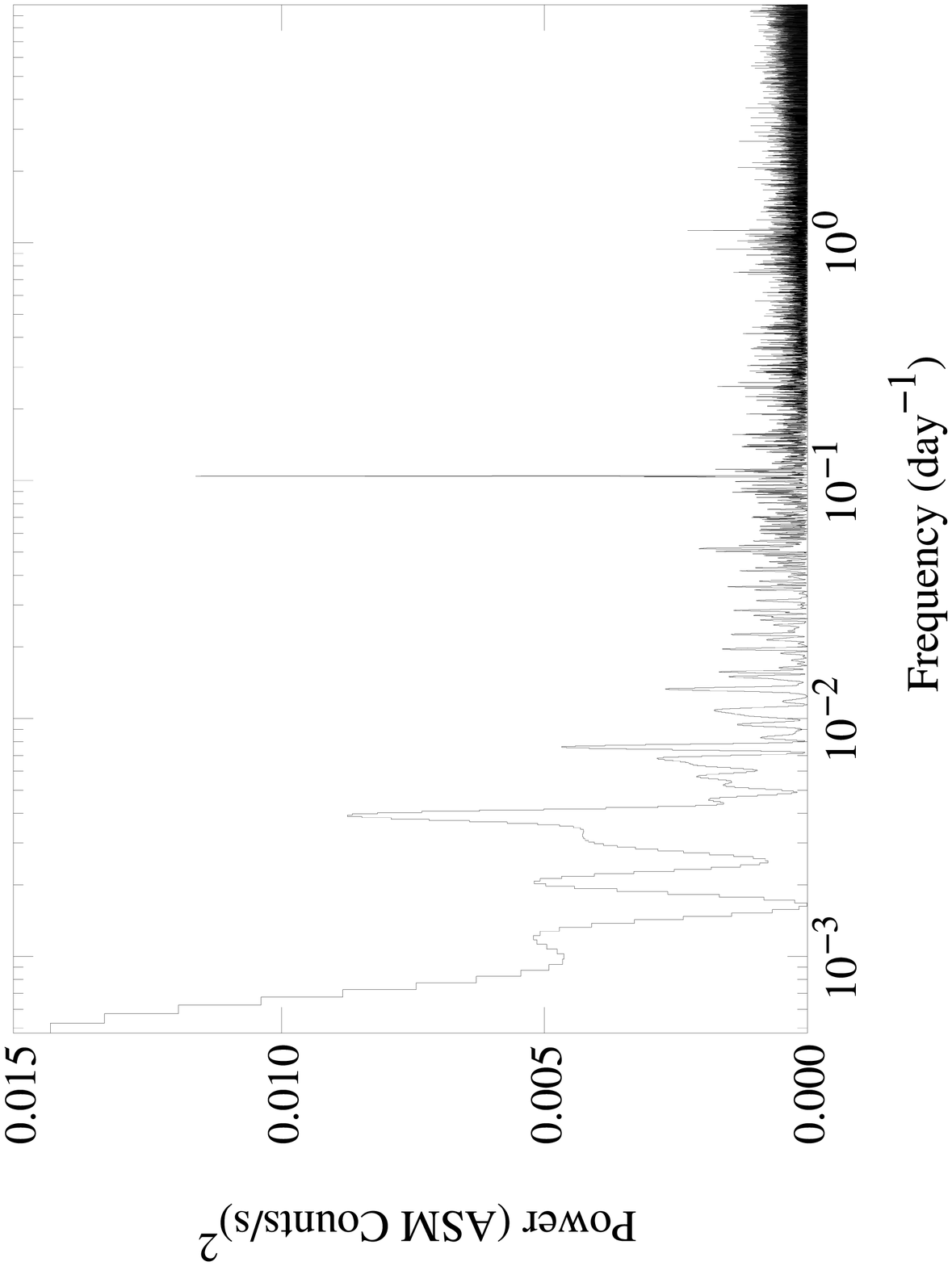]{Power spectrum of the ASM light curve
of \src. In addition to the periodic modulation
at \sqig10 days there is also some low frequency noise.}

\figcaption[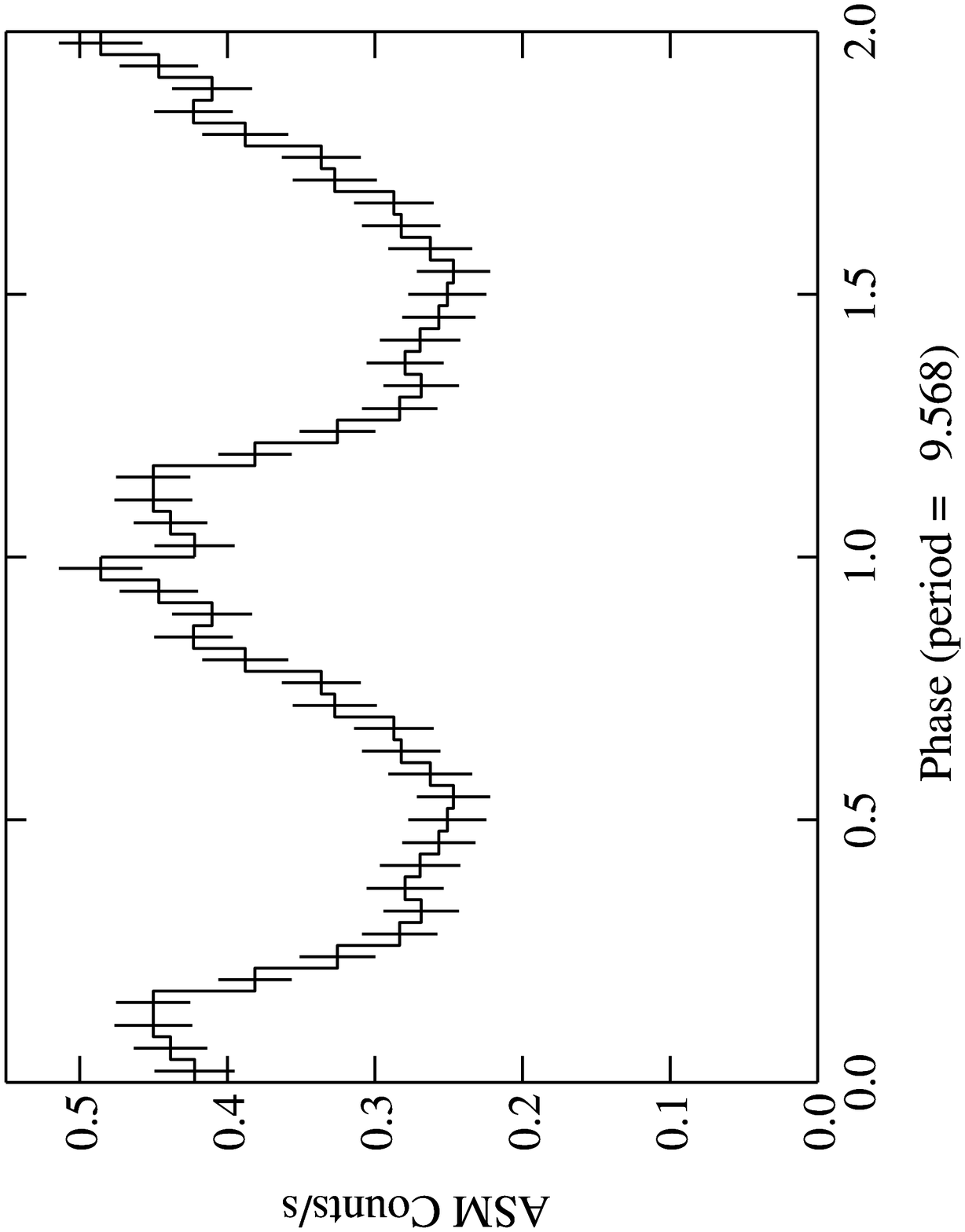]{RXTE ASM light curve of \src\ (individual
dwells) folded on the proposed
orbital period. 
The error bar for each bin is calculated using the standard
deviation and the number of points that contribute to that bin.}

\figcaption[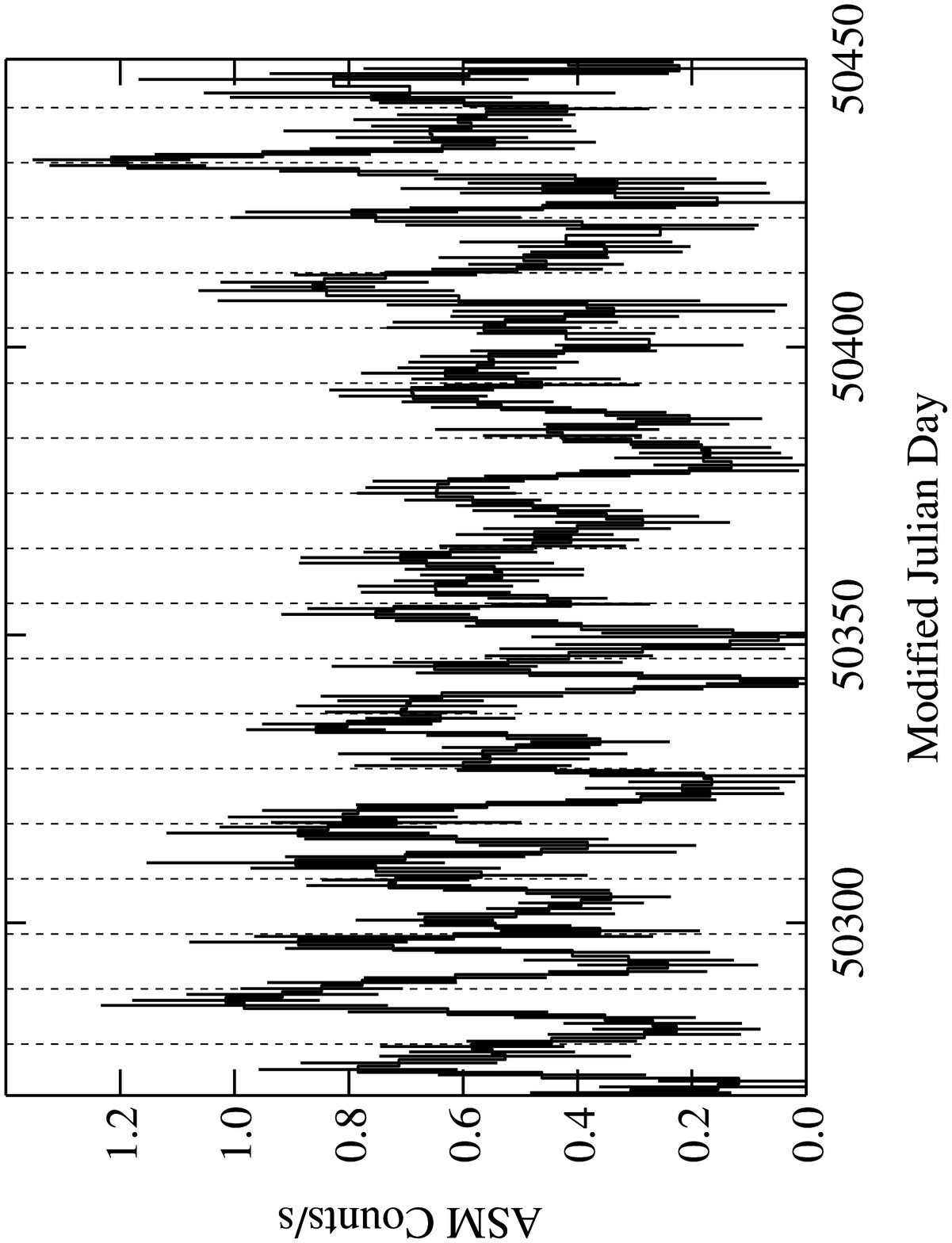]{A section of the ASM light curve of \src.
This is a smoothed version of the one day light curve but is not
rebinned as is Fig. 1. The
dashed lines indicate the expected maxima based on a sine wave fit
to the entire light curve.}

\figcaption[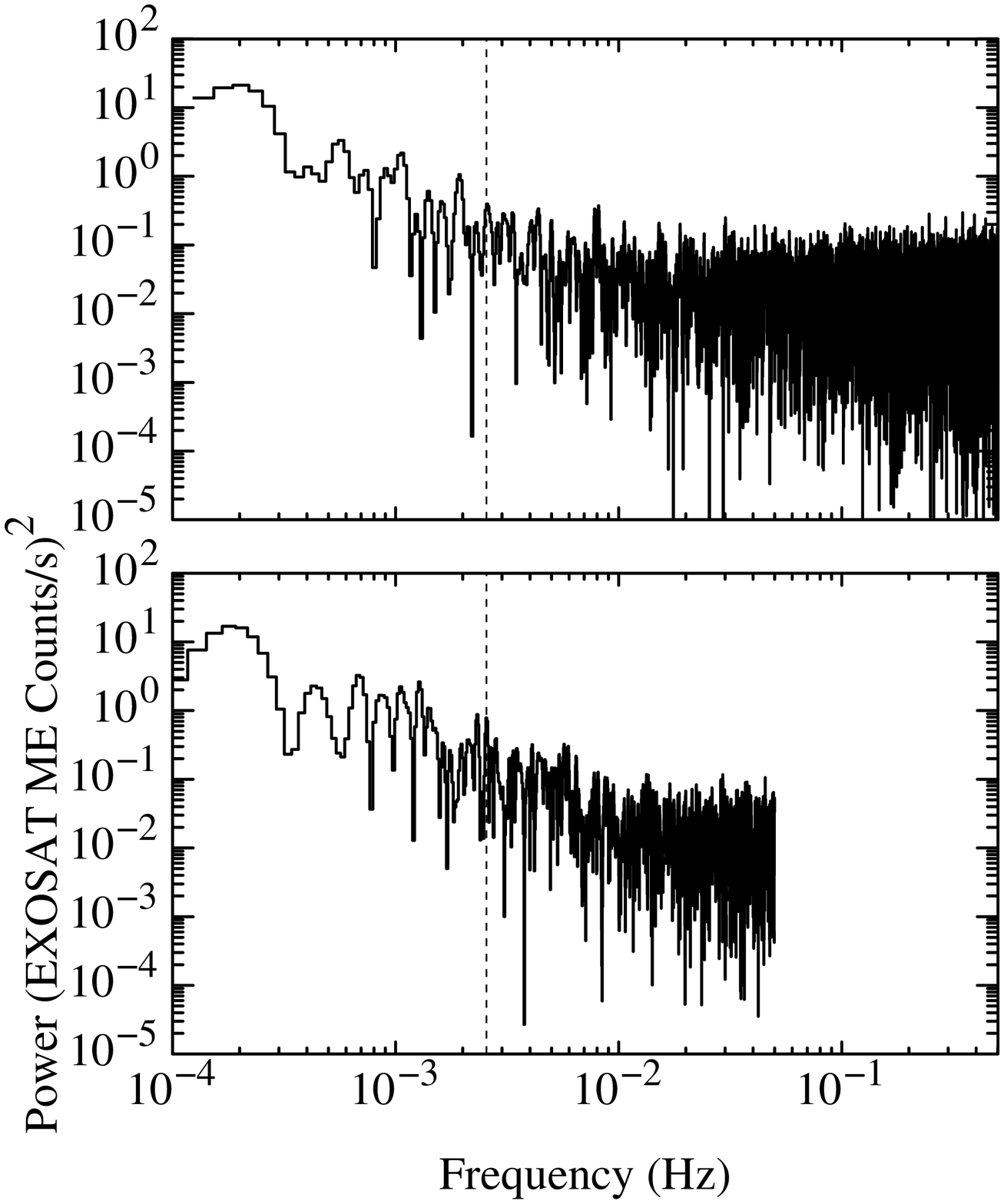]{(a) Bottom panel: The power spectrum
of the 10s time resolution light curve of \src\ obtained
with EXOSAT in 1983. 
(b) Top panel: The power spectrum
of the 1s time resolution light curve of \src\ obtained
with EXOSAT in 1985. The dashed line shows the frequency
corresponding to the 392s period reported by SA92.
The dashed lines show the frequency
corresponding to the 392s period reported by SA92.}

\figcaption[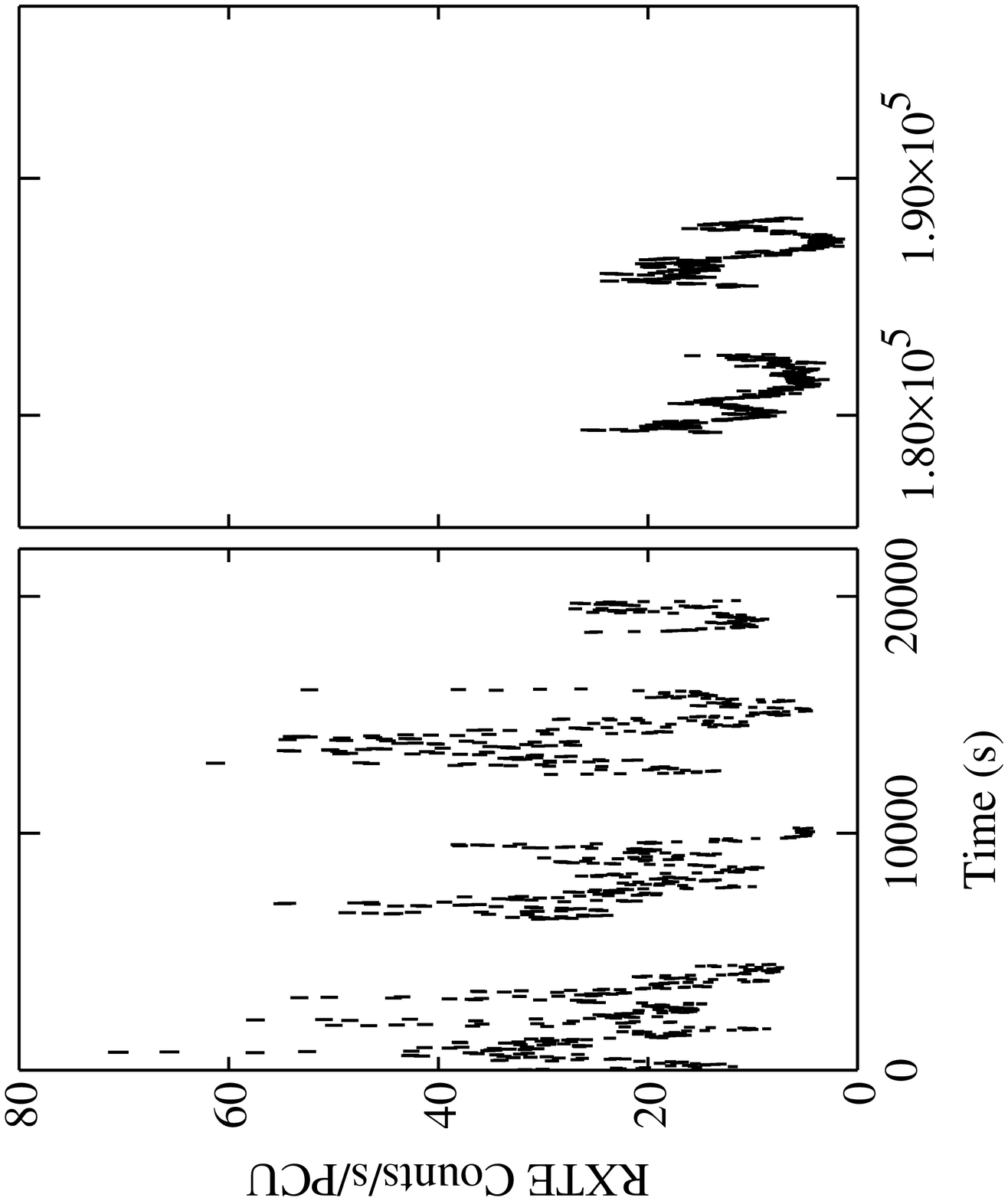]{Background subtracted light curve of \src\
obtained with the PCA detector on board RXTE. All five of the PCUs that
comprise the PCA were operating during the first observation and three
were operating during the second observation. Time is seconds since 
1997 March 11 4:13}

\figcaption[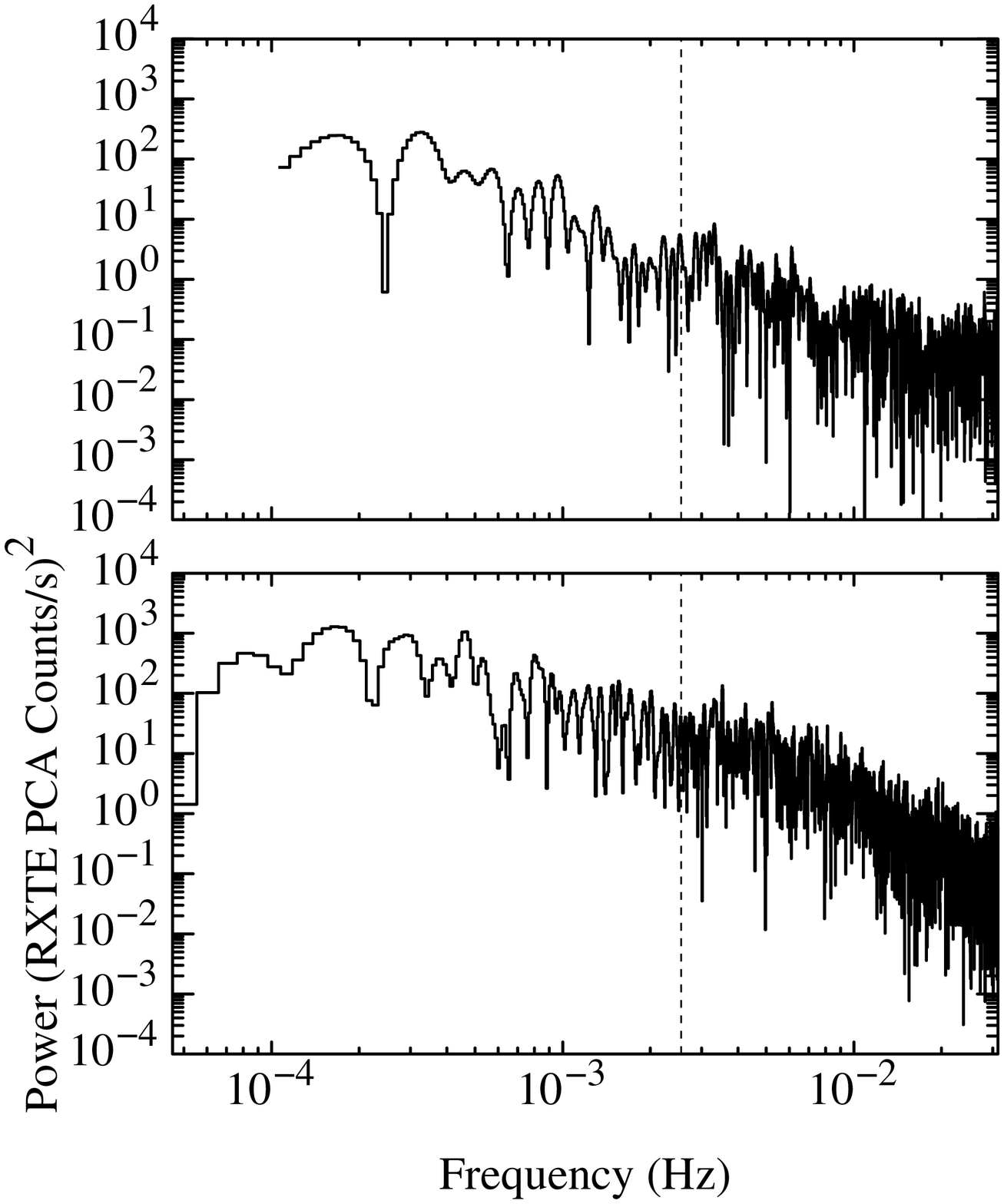]{The power spectra of the light curves of \src\
obtained with the RXTE PCA. (a) Bottom panel: power spectrum from the
observation performed on 1997 March 11. (b) Top panel: power spectrum
from the observation performed on 1997 March 13. The dashed lines show
the frequency corresponding to the 392s period reported by SA92.}


\begin{figure}
\plotone{f1.eps}
\end{figure}


\begin{figure}
\plotone{f2bit.eps}
\end{figure}


\begin{figure}
\plotone{f3.eps}
\end{figure}


\begin{figure}
\plotone{f4.eps}
\end{figure}

\begin{figure}
\plotone{f5.eps}
\end{figure}

\begin{figure}
\plotone{f6.eps}
\end{figure}

\begin{figure}
\plotone{f7.eps}
\end{figure}

\begin{table}
\begin{center}
\caption{Results of Fits to RXTE PCA Spectra}
\begin{tabular}{ccccccc}
\tableline
\tableline
Observation & $\alpha$ & N$_H$ & E$_{cut}$ & E$_{fold}$& $\chi^2_\nu$ & 2--10 keV Flux  \\
Number & &  ($\times$ 10$^{22}$) & & &  & (ergs cm$^{-2}$ s$^{-1}$)\\
\tableline
1 & 1.71 $\pm$ 0.03 & 4.6 $\pm$ 0.2 & 7.3 $\pm$ 0.1 & 17.3 $\pm$ 0.6 & 0.82 & 3.12 $\times$ 10$^{-10}$\\
2 & 1.12 $\pm$ 0.12 & 2.7 $\pm$ 0.7 & 5.3 $\pm$ 0.23 & 10.5 $\pm$ 1.2 & 0.75 & 1.14 $\times$ 10$^{-10}$\\
\tableline
\end{tabular}
\end{center}
Notes: The flux is the unabsorbed flux from the model fits.
\end{table}

\end{document}